\documentclass[9pt,letterpaper,twocolumn]{article}

\RequirePackage[utf8]{inputenc}
\RequirePackage[english]{babel}
\RequirePackage{amsmath,amsfonts,amssymb}
\RequirePackage{mathpazo} 
\RequirePackage[scaled]{helvet}
\RequirePackage[T1]{fontenc}
\RequirePackage{url}
\RequirePackage[colorlinks=true, allcolors=blue]{hyperref}

\RequirePackage{authblk}
\RequirePackage{graphicx,xcolor}
\definecolor{color1}{RGB}{59,90,198} 

\RequirePackage[left=48pt,%
                right=42pt,%
                top=46pt,%
                bottom=60pt,%
                headheight=15pt,%
                headsep=10pt,%
                letterpaper,twoside]{geometry}%

\usepackage[symbol]{footmisc}
\usepackage{graphicx}
\usepackage{color}
\usepackage{amssymb}
\usepackage{amsmath}

\RequirePackage[labelfont={bf,sf},%
                labelsep=period,%
                figurename=Fig.,%
                singlelinecheck=off,%
                font=footnotesize,%
               ]{caption}
\def\EQ#1{Eq.~(\ref{#1})}
\def\EQtwo#1#2{Eqs.~(\ref{#1}) and (\ref{#2})}

\def\Fig#1{Fig.\!\,\,\ref{#1}}

\def\Sec#1{Sec.~\ref{#1}}

\def\Vr{\Vec{r}}%
\def\Vs{\Vec{s}}%
\def\Vk{\Vec{k}}%

\newcommand\Fourier{{\mathcal F}}

\newcommand\FT[1]{\Fourier\!\{#1\}}
\newcommand\FTinvbig[1]{\Fourier^{-1}\!\left\{#1\right\}}
\newcommand\FTinv[1]{\Fourier^{-1}\!\{#1\}}

\def\Define{\triangleq}

\def\Abs#1{\left|#1\right|}

\def\Vec#1{{\boldsymbol{#1}}}

\def\Null{\rule{0pt}{0pt}}

\def\Obs{{\rm w}}
\def\ObsTr{\widehat{\rm w}}

\def\IpOpSym{\cal {I}}
\def\IpPolarOpSym{\widehat{\IpOpSym}}
\def\IpAveOpSym{\overline{\IpOpSym}}
\def\NatOpSym{{\cal N}}

\newcommand\Interpolate[1]{\IpPolarOpSym\left\{#1\right\}}

\newcommand\InterpolateAve[1]{\IpAveOpSym\left\{#1\right\}}
\newcommand\NatExt[1]{\NatOpSym\!\left\{#1\right\}}

\def\NIrep{{\NatOpSym\!,\,\IpOpSym}}
\def\NI{{\NatOpSym\!,\,\IpPolarOpSym}}
\def\NIave{{\NatOpSym\!,\,\IpAveOpSym}}

\def\Phiw{\Phi_{\Obs}}

\def\fw{f_{\Obs}}
\def\Fw{F_{\Obs}}

\def\Ext{{\rm ex\!}}
\def\fext{f_{\Ext}}
\def\wext{w_{\Ext}}
\def\what{\widehat{w}}
\def\Lhat{\widehat{L}}

\def\diff#1{d\left\{#1\right\}}

\def\Unwrapped#1{#1^{\rm u}}
\def\Max#1{\max\{#1\}}
\def\Min#1{\min\{#1\}}

\def\NondepOrigin#1{#1^{(0)}}

\def\ShiftOrigin#1{#1^{\dagger}}
\def\PhiShift{\ShiftOrigin{\Phi}}
\def\PhiShiftw{\ShiftOrigin{\Phiw}}
\def\thetaShift{\ShiftOrigin{\theta}}
\def\xshift{x_{\rm sft}}

\def\PhiSub#1{\Phi_{#1}}
\def\PhiIpPolarSub#1{\Phi_{#1}^{\IpPolarOpSym}}
\def\PhiIpPolarShiftSub#1{\Phi^{\dagger{\IpPolarOpSym}}_{#1}}

\def\OmegaObs{\Omega_{\Obs}}
\def\OmegaObsTr{\Omega_{\ObsTr}}
\def\OmegaExt{\Omega_{\Ext}}

\def\AveObs#1{\left\langle #1 \right\rangle_{\Obs}}

\def\AC#1{\widetilde{#1}}
\def\DC#1{\overline{#1}}

\newlength\HsizeInTwocol
\begin{document}
\twocolumn[

\noindent
\begin{flushleft}
{\huge\bf
Wavefront restoration from lateral shearing data\\
using spectral interpolation
}%
\end{flushleft}

\begin{center}
{%
\large\sc
  Satoshi Tomioka,$^{1,*}$
  Naoki Miyamoto,$^{1}$
  Yuji Yamauchi,$^{1}$
  Yutaka Matsumoto,$^{1}$\\
    and
  Samia Heshmat$^{2}$
}
\end{center}
\qquad
\parbox{0.9\hsize}{\small
  $^{1}$\ {\it Faculty of Engineering, Hokkaido University, Sapporo 060-8628, Japan}\\
  $^{2}$\ {\it Faculty of Engineering, Aswan University, Aswan 81542, Egypt}\\
  $^{*}$\ {\it Corresponding author: {\tt tom@qe.eng.hokudai.ac.jp}}
  \\
}

\noindent
\hfill
\parbox{0.9\hsize}{\Null\\%
\hrule
\vspace*{1em}%
\bf
Although a lateral-shear interferometer is robust against optical component vibrations,
its interferogram provides information about differential wavefronts
rather than the wavefronts themselves,
resulting in the loss of specific frequency components.
Previous studies have addressed this limitation by measuring
four interferograms with different shear amounts
to accurately restore the two-dimensional wavefront.
This study proposes
a technique that employs spectral interpolation to reduce the number of required interferograms.
The proposed approach introduces an origin-shift technique for accurate spectral interpolation,
which in turn is implemented by combining two methods:
natural extension and least-squares determination of ambiguities in uniform biases.
Numerical simulations confirmed that the proposed method
accurately restored a two-dimensional wavefront from just two interferograms,
thereby indicating its potential 
to address the limitations of the lateral-shear interferometer.
\\
\hrule
\vspace*{2em}%
}\hfill\Null
]
\footnotetext[0]{%
This manuscript is the author-created manuscript,
the contents of which were accepted for publication in Applied Optics
by Optica Publishing Group 
as a Research Article
on September 28, 2023.
The final Publisher Version of Record has a DOI of
\url{https://doi.org/10.1364/AO.500453}
with ``
\copyright 2023 Optica Publishing Group.
One print or electronic copy may be made for personal use only.
Systematic reproduction and distribution, duplication of any material in this paper for a fee or for commercial purposes, or
modifications of the content of this paper are prohibited.''
}

\section{Introduction}
\label{sec:intro}
A conventional interferometer utilizes two waves of light:
an object wave that passes through the object under investigation and a reference wave that does not interact with the object.
These waves must be nearly parallel to each other
to generate a fringe pattern in an interferogram.
However, 
when the angle of incidence on the object needs to be altered,
such as in computed tomography,
to measure the three-dimensional distribution of refractive index \cite{Tomioka:17}
using a mechanical stage,
it becomes difficult to
detect clear fringe patterns owing to the independent vibrations of individual optical components.

In contrast, a lateral shear interferometer (LSI) employs a sheared object wave instead of a reference wave, making it less susceptible to vibrations,
considering both the sheared and original object waves experience common vibrations.
However, the fringe analysis needed to obtain the wavefront in an LSI is more complex than that in a conventional interferometer.

Murty \cite{Murty:64} proposed 
a simplified configuration to generate the sheared wave, 
wherein the object wave was obliquely incident on a parallel glass plate,
resulting in two reflected waves at both the front and back surfaces.
The resulting two-dimensional (2-D) interfered fringe pattern is represented as
\begin{align}
  \label{eq:shear-interferogram}
  &I(\Vr)=I_0(\Vr)\cos\left(\Vk\cdot\Vr+f(\Vr)\right),
   \\
  \label{eq:shearing problem}
  &f(\Vr)=\phi(\Vr+\Vs)-\phi(\Vr),
\end{align}
where $\Vr$ is the position of the point observed on the screen,
$\Vk$ is the wavenumber vector of
the background fringe caused by the divergence of the incident beam and the angle of the glass plate,
$\phi$ is the wavefront to be restored,
and $\Vs$ is the shear amount.
Similar to the fringe analysis in conventional interferometry,
we can obtain the phase difference $f(\Vr)$
between the sheared and object waves
by applying background-fringe rejection using a Fourier transform \cite{Takeda:82}
and employing a phase-unwrapping method
\cite{Ghiglia:book:98,Takeda:96,Tomioka:10,Tomioka:12,Hirose:18,Heshmat:22}.
To restore $\phi(\Vr)$ from $f(\Vr)$, we must solve \EQ{eq:shearing problem}, called the shearing problem.

Over the past five decades, several methods have been proposed
to address the mathematical complexities associated with the shearing problem.
These methods can be broadly classified into two categories: modal methods and zonal methods.

In modal methods, the wavefront $\phi(\Vr)$ is represented by a polynomial function, such as the Zernike polynomial.
The coefficients of the basis functions are estimated using least-squares fitting 
\cite{Rimmer:75,Harbers:96,Okuda:00,Dai:13,Mochi:15,Ling:15}.
These methods impose a limitation on the object being investigated to ensure
the wavefront can be adequately represented by the polynomial function.

Conversely, zonal methods directly estimate the wavefront at the grid points.
However, these methods encounter rank deficiency issues.
For example,
consider a one-dimensional (1-D) problem where
the wavefront difference $f(x)$ is sampled at $N$ points at intervals of $\Delta x$ as
\begin{align}
  \label{eq:f_i}
  f^x_{i}=\phi_{i+n_s}-\phi_{i}  \quad (i=\{0,1,\cdots,N-1\}),
\end{align}
where $n_s$ is the number of points for the shear amount ($n_s=s/\Delta x$).
In the case of $n_s=1$, \EQ{eq:f_i} can be arranged as
\begin{align}
  \phi_n&=\phi_{n-1}+f^x_{n-1}=\phi_{n-2}+f^x_{n-2}+f^x_{n-1}=\cdots
         \nonumber\\&
         =\phi_0+\sum_{i=0}^n f^x_i.
\end{align}
Considering the right-hand side comprises known values except for $\phi_0$,
we can successively restore $\phi_n$ on the left-hand side
if $\phi_0$ is predefined.
In the context of wavefront restoration, $\phi_0$ represents a uniform offset of every $\phi_i$
which is not crucial for the restored wavefront.
Considering $f^x(x)$ as the derivative of $\phi(x)$
and $\phi(x)$ as the integral of $f^x(x)$,
$\phi_0$ corresponds to an integral constant
known as `piston term' or `bias term.'
Similarly, in the case of $n_s\ge 2$, $\phi$ can be represented as
\begin{align}
  \label{eq:ns_shear}
  \phi_{n+m}&=\phi_m+\sum_{j=0}^{\lfloor n/n_s\rfloor} f^x_{j n_s+m}
   & \quad (m=\{0,1,\cdots,n_s-1\}).
\end{align}
On rearranging \EQ{eq:ns_shear},
$n_s$ unknown variables remain, which indicates that this problem is $n_s$-rank deficient.
If we assume $n_s$ additional equations, such as $\phi_m=0$ for $i=\{0,\cdots,n_s-1\}$,
or if the shear amount exceeds the object size by a significant margin
\cite{TAYAL2019991,tang2021retrieval},
we can solve
all of $\phi_{i'}$ for $i'=\{0,\cdots,N+n_s-1\}$.
However, this technique is not universally applicable.

In the case of a 2-D problem, $f^x_{i,j}=\phi_{i+n_s,j}-\phi_{i,j}$ for $i,j=\{0,\cdots,N-1\}$,
the restored result includes $n_s$ unknown variables for each $j$,
which is the $N n_s$-rank deficient problem.
Using additional measurements with different shear directions, such as $f^y_{i,j}$ in addition to $f^x_{i,j}$,
we can increase the number of equations.
This effectively doubles the number of equations without increasing the number of variables,
which generally indicates an overdetermined problem.
However, previous studies
\cite{Hudgin:77,Fried:77,Frost:79,Hunt:79,Southwell:80,Herrmann:80,Legarda-Saenz:00,Zou:00,Zhai:09,Dai:12,Bloemhof:14}
show that this problem cannot be solved using the least-squares method.
Additionally, the set of normal equations for the least-squares problem remains rank deficient
\cite{Herrmann:80,Southwell:80,Zou:00,Dai:12,Bloemhof:14}.
As a result, the wavefront is obtained as a minimum norm solution,
which is one of the possible solutions,
rather than the least-squares solution.
To resolve the rank-deficient problem,
a multi-shear restoration approach \cite{DAI2016264,Zhai:16,Zhai:17}
that uses four interferograms with two different shears for each $x$ and $y$ direction
has been proposed.
However, the simultaneous measurement of
multiple interferograms introduces complexity to the measurement system
and requires specialized techniques.

Restoration methods utilizing Fourier transform
\cite{Hudgin:77,Freischlad:86,Elster:99,Elster:99-B,Elster:00,Dubra:04,YIN2005456,Liang:06,Guo:12,Guo:14,Zhai:17-high,ling2017quadriwave}
are also classified as zonal methods.
The Fourier transform assumes implicit periodicity,
where the data within a finite domain (with a domain size of $L=N\Delta x$) is repeated outside the original domain,
such as $\phi(x)=\phi(x+L)$ in the case of 1-D.
This periodicity, which can be viewed as additional conditions, 
prevents rank deficiency issues in methods employing the Fourier transform.
However, if $f(x)$ does not converge to the same value at the domain ends,
the periodicity assumption becomes inappropriate, considering $f(x)$ for $x\in[L-s,L)$ is treated as
$f(x)=\phi(x+s-L)-\phi(x)\ne\phi(x+s)-\phi(x)$.
This problem is known as a non-periodic or limited window problem.
Elster and Weing\"{a}rtner \cite{Elster:99} demonstrated that
the non-periodic problem in a 1-D shearing problem
can be addressed using the `natural extension' method,
which involves adding data that satisfies a specific condition.
Another issue that arises in Fourier transform-based restoration
is that the wavefront difference $f(x)$ does not convey information about the spectral components of $\phi(x)$ of a frequency of $1/s$ and its integer multiples.
This can be illustrated by the simple example of $\phi(x)=\cos(2\pi x/s)$.
Here, $\phi(x)$ satisfies $\phi(x+s)=\phi(x)$; therefore, $f(x)=0$.
To tackle this problem,
methods using two interferograms with different shears for each direction
have been proposed \cite{Elster:99,Elster:99-B,Elster:00,Dubra:04,Guo:12,Zhai:17-high,ling2017quadriwave},
similar to zonal methods, which employ the least-squares method in the spatial domain.
When the two shear amounts in each direction are coprime and combined with natural extension \cite{Elster:99-B, Elster:00},
exact 2-D solutions can be restored.
However, fine-tuning shear amounts in addition to multiple shear measurements would require significant experimental efforts.

Other methods have been proposed to recover the missing frequency components,
including spectral interpolations 
using simple averaging \cite{Liang:06} and resampling \cite{Falldorf:07} techniques.
These methods do not require
additional difference data with varying shear amounts.
However, for accurate interpolations,
it is necessary for the neighboring spectrum around the frequency to be interpolated
to exhibit smooth behavior.
When the lost frequencies are located on a spectral tail
outside the main lobe of the wavefront spectrum,
the intensity of the spectrum remains smooth;
however, rapid phase changes can occur in some cases.
To address this issue, this study proposes a novel technique called origin shift,
where the origin shift interpolation is applied to the Fourier spectrum of
the 1-D difference.

Coupling the origin-shift interpolation method with natural extension
can effectively restore wavefronts with minimal error.
However, this method is primarily applicable to 1-D problems.
To restore a 2-D wavefront, it is necessary to address the piston term problem.
By utilizing a single 2-D difference data set $f^x_{i,j}$, wherein the shear is in the $x$ direction,
$\phi^{x}_{i,j}$ can be evaluated for each $j$
with piston terms, $\overline{\phi^x_j}$, which cannot be determined uniquely.
Similarly, the difference function $f^y_{i,j}$ measured with $y$-directional shear,
$\phi^{y}_{i,j}$ has piston terms of $\overline{\phi^y_i}$.
These piston terms can be determined
using the least-squares method proposed by Tian {\it et al}. \cite{Tian:95}.

The objective of this study is to precisely restore a 2-D wavefront
using only two 2-D differences acquired from different shear directions.
It is important to note that
each difference does not converge to the same value, which indicates a non-periodic problem.
To ensure accurate reconstruction of 2-D wavefronts, we employ a combination of three methods:
the natural extension method for addressing the non-periodic nature of the problem,
proposed spectral interpolation with origin shift,
and determining the piston terms using the least-squares method.

The remainder of this paper is organized as follows.
Section \ref{sec:theory} provides a comprehensive review of
the restoration techniques involving the Fourier transform and natural extension,
along with proof of 
why the natural extension yields exact results.
Then, we introduce
the spectral interpolation approach with the origin shift.
To evaluate the effectiveness of our method,
we present
numerical simulations
for 1-D and 2-D wavefront restorations
in \Sec{sec:result}.
Finally, \Sec{sec:conclusion} discusses
the key characteristics of our proposed method.

\section{Wavefront Restoration Algorithm}
\label{sec:theory}

In this section, we propose a method
for restoring a 1-D wavefront from difference data
by combining spectral interpolation with an origin shift technique 
in non-spectral space
to achieve accurate spectral interpolation.
The restoration is then enhanced using a natural-extension method.
First, we discuss the limitations of restoration techniques based on the Fourier transform.
Subsequently, we validate the correctness of the natural extension method.
Finally, we present a precise spectral interpolation method employing an origin shift.

\subsection{Shearing transfer function}
\label{subsec:shearing_transfer_function}

A 1-D wavefront difference $f(x)$ acquired by measurement
is defined as the difference between the wavefront $\phi(x)$ and a sheared wavefront $\phi(x+s)$ with the shear amount $s$
as
\begin{align}
  \label{eq:observe}
  f(x)=\phi(x+s)-\phi(x) \Define \diff{\phi(x)},
\end{align}
where $\diff{\ }$ represents a difference operator.
In the above equation, $f(x)$ and $\diff{\phi(x)}$ are also the functions of $s$; however, we omit the explicit notation $s$ for simplicity.

The Fourier transform of $f(x)$ can be expressed as the product of two functions:
a specific function called the `shear transfer function'
that corresponds to the Fourier transform of the operator $\diff{\ }$
and the Fourier transform of $\phi(x)$.

Before evaluating the derivation of this relationship, 
we provide definitions of the Fourier transform and
highlight its significant properties.

This study defines
a pair of forward and inverse Fourier transforms as
\begin{align}
  \label{eq:FT-def}
  \FT{a(x)}&\Define\int_{-\infty}^{\infty} a(x) e^{-2\pi i\kappa x}\,dx=A(\kappa),
  \\
  \label{eq:IFT-def}
  \FTinv{A(\kappa)}&\Define\int_{-\infty}^{\infty} A(\kappa) e^{+2\pi i\kappa x}\,d\kappa=a(x),
\end{align}
where $i$ is the complex unit
and $\kappa$ is the wavenumber (spatial frequency) divided by the dimension of $x$. 
When the data is sampled at intervals of $\Delta x$ within a finite range of $x\in\Omega=[x_0,x_0+L)$,
we obtain a discrete Fourier transform (DFT) by approximating
$\int\,(\cdots)\,dx\simeq \sum(\cdots)\Delta x$ as
\begin{align}
  \label{eq:DFT}
    A_m
      &=\sum_{n=0}^{N-1} a_n\,e^{-2\pi i\kappa_m x_n}\Delta x \qquad (m=\{m_{\min},\cdots,m_{\max}\}),
      \\
  \label{eq:IDFT}
    a_n
      &=\sum_{m=m_{\min}}^{m_{\max}} A_m \,e^{+2\pi i \kappa_m x_n}\Delta\kappa \qquad (n=\{0,\cdots,N-1\}).
\end{align}
In the above equations, $a_n$ and $A_m$ represent $a(x_n)$ and $A(\kappa_m)$, respectively,
and $N$ represents the number of samples ($N=L/\Delta x$).
To adhere to the well-known sampling theorem,
$m_{\min}$ and $m_{\max}$ are determined
using the floor and ceil functions
as $m_{\min}=-\lfloor{N/2}\rfloor$, $m_{\max}=\lceil{N/2}\rceil-1$, respectively.
These quantities satisfy the following relations:
$m_{\max}-m_{\min}=N-1$, $\kappa_m=m\Delta\kappa$, and $\Delta\kappa=1/L$.
Going forward, the values with dimensions are measured in the sampling interval,
particularly in pixels,
while spatial frequencies are measured in cycles per pixel.
Note that there exist periodicities
\begin{align}
  \label{eq:periodicity-real}
  & a_{n+lN}=a_{n},\\
  \label{eq:periodicity-spectrum}
  & A_{m+lN}=A_{m},
\end{align}
where $l$ represents any integer,
which introduces an error in the restoration of the wavefront $\phi(x)$ from
the difference $f(x)$ when employing the Fourier transform.
This will be discussed in detail later.

A Fourier transform of the sheared function $\phi(x+s)$ in the right-hand side of \EQ{eq:observe} can be evaluated as
\begin{align}
  \label{eq:shift_fourier}
  \FT{\phi(x+s)}
    &=\int_{-\infty}^{\infty} \phi(x+s) e^{-2\pi i\kappa x}\,dx
    \nonumber\\&
    =e^{2\pi i \kappa s}\int_{-\infty}^{\infty} \phi(x+s) e^{-2\pi i\kappa (x+s)} \,dx
    \nonumber\\&
    =e^{2\pi i\kappa s}\FT{\phi(x)}.
\end{align}
Using this relation and \EQ{eq:observe}, a Fourier transform of the observed difference function, $f(x)$, can be given as
\begin{align}
   \label{eq:F_from_Phi}
   F(\kappa)&\Define \FT{f(x)}=D(\kappa)\FT{\phi(x)},
   \\
   \label{eq:D-def}
   D(\kappa)&=e^{2\pi i \kappa s}-1,
\end{align}
where $D(\kappa)$ is the shear transfer function.
In \EQ{eq:F_from_Phi}, $F(\kappa)$ can be computed from the observed function $f(x)$;
therefore, $\phi(x)$ can be evaluated in principle as
\begin{align}
  \label{eq:restore_using_shearing_transfer}
  \phi(x)&=\FTinvbig{\frac{F(\kappa)}{D(\kappa)}}.
\end{align}

\begin{figure}[t]
{%
 \hfill
 \parbox{\HsizeInTwocol}{%
  {\hfill\scalebox{0.9}{\parbox{\hsize}{%
    \parbox{\hsize}{(a)\\
      {\hfill
        \includegraphics[width=\hsize]{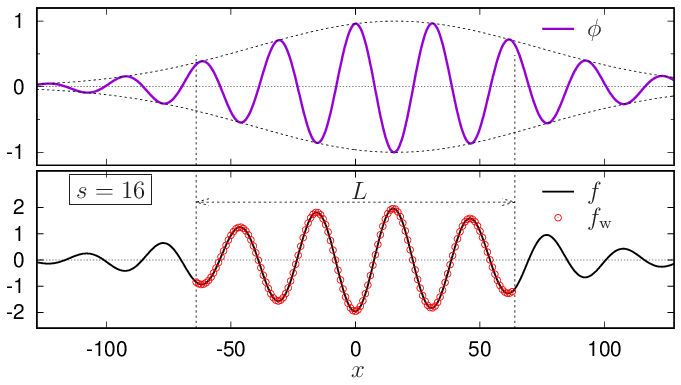}
      \hfill\Null}
    }%
    \\[-1em]%
    \parbox{\hsize}{(b)\\
      {\hfill
        \includegraphics[width=\hsize]{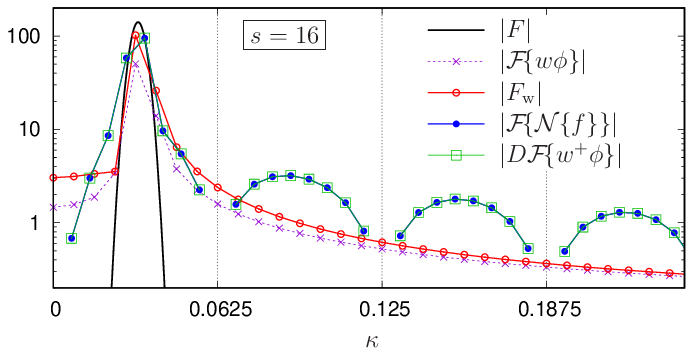}
      \hfill\Null}
    }%
  }%
  }\hfill\Null}%
 }%
 \hfill\Null
  \caption{
    Effect of the limited window on spectrum distortion:
    (a) Original wavefront $\phi(x)$
        and corresponding wavefront difference $f(x)$ for $s$=16, 
        with the observed data in the limited range $\fw(x)$.
        The range width is [-64,64) 
        and the sampling interval is $\Delta x$=1 ($N$=128, $L$=128).
        Wavefront parameters are $(\lambda,x_c,w_g)$=(31,16,80) for
        ${\phi}(x)=\cos(2\pi x/\lambda)e^{-(x-x_c)^2/w_g^2}$.
    (b) Spectral intensities related to the functions of $f(x)$, $\phi(x)$.
        Vertical dashed lines indicate that $\kappa$ is a multiple of $1/s$.
  }
  \label{fig:test_sig1}
}%
\end{figure}

Equation (\ref{eq:restore_using_shearing_transfer}) 
highlights a fundamental issue that is encountered by employing the Fourier transform for restoration purposes.
This issue arises
at $\kappa=1/s\Define\kappa_s$ and its integer multiples,
called `shear frequencies.'
The restoration of $\phi(x)$ encounters fails at shear frequencies
owing to the presence of $D(\kappa)$ in the denominator of the right-hand side, resulting in a division by zero.
It is important to note that both the denominator $D(\kappa)$ and numerator $F(\kappa)$ become zero
at shear frequencies
considering the numerator $F(\kappa)$ is lost during the measurement process, as explained in \Sec{sec:intro}.
If we were to approach the shear frequencies by taking the limit of $\kappa$, we can determine $F(\kappa)/D(\kappa)$ for all $\kappa$
and subsequently restore $\phi(x)$ using an inverse Fourier transform.
The methodology illustrating this approach will be presented in detail in Sec.~\ref{subsec:interpolation}.

This issue can be mitigated by adjusting $s$ to eliminate the components at the shear frequencies;
i.e., $\Phi(\kappa)\simeq 0$ at $|\kappa|\ge|\kappa_s|$,
which in turn we can establish $F(\kappa)/D(\kappa)=0$ at those frequencies and achieve restoration.
However,
when the observed domain is limited to finite size and the periodicity of $\phi(x)$ indicated in \EQ{eq:periodicity-real} is not satisfied,
 significant restoration errors appear even after adjusting $s$.
Owing to the non-periodicity, a spectral tail known as `leakage' extends into the higher frequency region,
thereby causing the wavefront spectrum for the finite domain to exhibit higher intensities than that of an infinite domain.
Therefore, $\Phi(\kappa)$ cannot be considered zero for the limited window.
One possible approach to address the non-periodic problem is elucidated in the subsequent subsection.

\subsection{Natural Extension}
\label{subsec:natural extension}
In practical measurements,
the measurement domain $\OmegaObs$ is constrained to a finite range
whereas the domain of the difference of $f(x)$ can be considered infinite.
The observed difference is represented by multiplying it with a window function $w(x)$, as $w(x)f(x)$, where
\begin{align}
  w(x)&=\left\{\begin{array}{ll}
    1 & \mbox{for } x\in\OmegaObs\\
    0 & \mbox{for } x\not\in\OmegaObs
  \end{array}\right. 
  .
\end{align}
The Fourier spectrum of the finite window $\Fw(\kappa)$ is determined using \EQtwo{eq:shift_fourier}{eq:D-def}, yielding
\begin{align}
  \label{eq:FT-wf}
  \Fw(\kappa)\Define &\FT{w(x)f(x)}
    \nonumber\\
    =&\FT{w(x)\phi(x+s)-w(x)\phi(x)}
    \nonumber\\
    =&\FT{w(x+s)\phi(x+s)-w(x)\phi(x)}
    \nonumber\\
     &+\FT{(w(x)-w(x+s))\phi(x+s)}
    \nonumber\\
    =&D(\kappa)\FT{w(x)\phi(x)}
     +\Delta\Phiw(\kappa)
    ,
  \\
  \label{eq:Def-DeltaPhiw}
  \Delta\Phiw(\kappa)&=\FT{(w(x)-w(x+s))\phi(x+s)}
  .
\end{align}
Figure \ref{fig:test_sig1} shows an example of $\Fw(\kappa)$.
Here, $\phi(x)$ shown in (a) is defined
as a product of 
a Gaussian function and a sine function; therefore,  $|\FT{\phi(x)}|$ shows a Gaussian function.
However, owing to the shear frequency $\kappa_s$ exceeding the dominant spectral component of $\phi(x)$,
no dip is observed in $|F(\kappa)|$, as denoted by the black line in (b).
Within the limited domain, $|\FT{w(x)\phi(x)}|$ exhibits a tail due to spectral leakage (denoted by the purple dashed line)
resulting from that $w(x)\phi(x)$ does not satisfy the condition for periodicity.
Despite the expectation for dips at the shear frequencies
when $D(\kappa)$ is multiplied by $\FT{w(x)\phi(x)}$,
they are absent
in $|F_w(\kappa)|$ (denoted by the red line) 
owing to the additional term $\Delta\Phiw(\kappa)$ presented in \EQ{eq:FT-wf}.
If $\Delta\Phiw(\kappa)=0$ were to be applied, $w(x)\phi(x)$ could be obtained from $\Fw(\kappa)$
through the restoration process outlined in \EQ{eq:restore_using_shearing_transfer}.
In the Fourier kernel of $\Delta\Phiw(\kappa)$,
$\phi(x+s)$ is multiplied by $w(x)-w(x+s)$, which possesses a non-zero value near the ends of $\OmegaObs$.
Unless the values of $\phi(x)$ at the ends coincide (satisfying the condition for periodicity), the kernel remains non-zero.
This is the reason for $\Delta\Phiw(\kappa)\ne0$.

To cancel $\Delta\Phiw(\kappa)$,
Elster and Weing\"{a}rtner \cite{Elster:99} proposed a method called
`natural extension.'
However, it should be noted that the representation provided below diverges from that of the original paper,
because the definition of $f(x)$ has been modified.

A natural extension technique is applied when the domain size of the observed data $L$
is an integer multiple of $s$.
Else, $L$ is adjusted
using either a smooth extension \cite{Elster:99} or truncation \cite{Elster:99-B} before applying this technique.
This study explored the characteristics of the natural extension when the data size is truncated to $\Lhat=\lfloor L/s\rfloor s$.
The appended data resulting from the natural extension are determined as follows, considering the observed wavefront difference $f(x)$ defined by \EQ{eq:observe}:
\begin{align}
  \label{eq:Nat-ext}
  \fext(x)&=-\sum_{p=1}^{\Lhat/s} f(x+ps) \quad
       (x\in\OmegaExt).
\end{align}
Here, $\OmegaExt$ represents the domain where the additional data in the width $s$ is appended,
specifically on the negative side of $\OmegaObsTr=[x_0,x_0+\Lhat)$
(i.e., $\OmegaExt=[x_0-s,x_0)$).
Notably, we chose the negative side for ease of proof;
however, the positive side of $\OmegaObsTr$
can also be selected owing to the periodic condition
of the Fourier transform, as shown in \EQ{eq:periodicity-real}.
The data obtained after applying the natural extension can be represented as
\begin{align}
  \label{eq:natural_extension}
  &\NatExt{f(x)}=\what(x)f(x)+\wext(x)\fext(x)
  ,
    \\
  &\what(x)=\left\{%
     \begin{array}{ll}%
       1 & \mbox{for } x\in\OmegaObsTr\cr%
       0 & \mbox{for } x\not\in\OmegaObsTr%
     \end{array}%
     \right.
  \!\!\!\!,\ 
  \wext(x)=\left\{%
     \begin{array}{ll}
       1 & \mbox{for } x\in\OmegaExt\cr
       0 & \mbox{for } x\not\in\OmegaExt
     \end{array}%
     \right.
  \!\!\!\!.
\end{align}
The support function of observed window $\what(x)$ can be expressed
as the sum of shifted functions of $\wext(x)$, represented as
\begin{align}
  \label{eq:w_hat_sum}
  \what(x)=\sum_{p=1}^{\Lhat/s} \wext(x-ps).
\end{align}

Before evaluating the Fourier transform of \EQ{eq:natural_extension},
the first term on the right-hand side
is rearranged using the similar procedure in the derivation of \EQ{eq:FT-wf},
yielding the modified \EQ{eq:natural_extension} as
\begin{align}
  \label{eq:natural_extension-2}
  \NatExt{f(x)}=&
      (\what(x+s)\phi(x+s)-\what(x)\phi(x))
    \nonumber\\&
      +(\what(x)-\what(x+s))\phi(x+s)
    \nonumber\\&
      +\wext(x)\fext(x).
\end{align}
The Fourier transform of the first term on the right-hand side can be evaluated as
\begin{align}
  \FT{\what(x+s)\phi(x+s)-\what(x)\phi(x)}
     =(D(\kappa)-1)\FT{\what(x)\phi(x)}.
\end{align}
The second term on the right-hand side in \EQ{eq:natural_extension-2}
is expressed as two sums of shifted functions of $\wext(x)$ using \EQ{eq:w_hat_sum},
and upon canceling common terms,
it is rewritten as 
\begin{align}
  \label{eq:diff_w.phi}
  \bigl(\what(x)&-\what(x+s)\bigr)\phi(x+s)
    \nonumber\\&
    =\sum_{p=1}^{\Lhat/s} \left(\wext(x-ps)-\wext(x-(p-1)s)\right)\phi(x+s)
    \nonumber\\&
    =\left(\sum_{p=1}^{\Lhat/s} \wext(x-ps)-\sum_{p'=0}^{{\Lhat/s}-1} \wext(x-p's)\right)\phi(x+s)
    \nonumber\\&
    =\wext(x-\Lhat)\phi(x+s)-\wext(x)\phi(x+s)
   .
\end{align}
Similarly, in the third term on the right-hand side of \EQ{eq:natural_extension-2}, $\wext(x)\fext(x)$ can be substituted with
\begin{align}
  \label{eq:w.diff_phi}
  \wext(x)\fext(x)
    &
    =-\wext(x)\sum_{p=1}^{\Lhat/s}\bigl(\phi(x+(p+1)s)-\phi(x+ps)\bigr)
    \nonumber\\&
    =-\wext(x)\phi(x+(\Lhat+s))+\wext(x)\phi(x+s)
   .
\end{align}
The right-hand sides of \EQtwo{eq:diff_w.phi}{eq:w.diff_phi}
contain common terms with opposite signs.
The Fourier transforms of the remaining non-common terms in \EQtwo{eq:diff_w.phi}{eq:w.diff_phi} are evaluated
using \EQ{eq:shift_fourier} as described in
\begin{align}
  \label{w(x-L)phi(x+s)}
     &
   \FT{\wext(x-\Lhat)\phi(x+s)}
    \nonumber\\&\hspace*{4em}
    =e^{2\pi i\kappa s}\FT{\wext(x-(\Lhat+s))\phi(x)},
  \\
  \label{w(x)phi(x+(L+s))}
     &
   \FT{-\wext(x)\phi(x+(\Lhat+s))}
    \nonumber\\&\hspace*{4em}%
    =-e^{2\pi i\kappa (\Lhat+s)}\FT{\wext(x-(\Lhat+s))\phi(x)}.
\end{align}
When the domain size is $\Lhat+s$, the frequency $\kappa$ in the DFT is sampled so that $\kappa(\Lhat+s)$ is an integer;
therefore,
the factor on the right-hand side of \EQ{w(x)phi(x+(L+s))} is evaluated as
\begin{align}
  e^{2\pi i\kappa (\Lhat+s)}=1.
\end{align}
Consequently, the DFT of the extended data presented in \EQ{eq:natural_extension} can be expressed as
\begin{align}
  \label{eq:nat-FT}
  \FT{\NatExt{f(x)}}&
   =(e^{2\pi i\kappa s}-1)\FT{\what(x)\phi(x)}
    \nonumber\\&
   +(e^{2\pi i\kappa s}-1)\FT{\wext(x-(\Lhat+s))\phi(x)}
    \nonumber\\&
   =D(\kappa)\FT{w^{+}(x)\phi(x)}
   ,\\
  \label{eq:w_Nat_right_pad}
  w^{+}(x)&\Define \what(x)+\wext(x-(\Lhat+s)),
\end{align}
where the second term on the right-hand side of \EQ{eq:w_Nat_right_pad} represents the support for the region located outside the right-hand side of $\OmegaObs$.

Equation (\ref{eq:nat-FT}) demonstrates that the Fourier transform of the extended difference by the natural extension technique
is equal to the product of the shear transfer function $D(\kappa)$ and the Fourier transform of $\phi(x)$ in the extended domain.
To illustrate this equality,
a numerical example
is presented in \Fig{fig:test_sig1}, represented by small blue-filled circles and open green squares.

After applying the restoration procedure to \EQ{eq:nat-FT}, we obtain the relationship as
\begin{align}
  \label{eq:FTinvPhiNat}
  \FTinvbig{\frac{\FT{\NatExt{f(x)}}}{D(\kappa)}}
   =\what(x)\phi(x)
   +\wext(x-(\Lhat+s))\phi(x).
\end{align}
Notably, $\what(x)$ and $\wext(x-(\Lhat+s))$, for any $x$, do not become unity simultaneously.
However, we can determine $\phi(x)$ at $x\in[x_0,x_0+\Lhat+s)$ by evaluating the left-hand side.
This relationship reveals an interesting characteristic.
The observed data domain, with a size of $L$, is truncated to $\Lhat$,
while the estimated wavefront has a wider size of $\Lhat+s$.
Because the number of data points is determined by dividing the domain size by the sampling interval,
the truncated data corresponding to the input of wavefront estimation
is fewer than
the estimated data corresponding to the output.
Despite having fewer input data points than the output,
the method can accurately estimate the wavefront without errors.
The unresolved issue lies in the treatment of $1/D(\kappa)$,
particularly with respect to the shear harmonic frequencies.

\subsection{Spectral Interpolation with Origin Shift}
\label{subsec:interpolation}
\begin{figure}[btp]
\hfill
\parbox{\HsizeInTwocol}{%
 \hfill
 \parbox{0.45\hsize}{%
  {\hfill\scalebox{1.0}{\parbox{\hsize}{%
    \parbox{\hsize}{(a)\\
      {\hfill
        \includegraphics[width=\hsize]{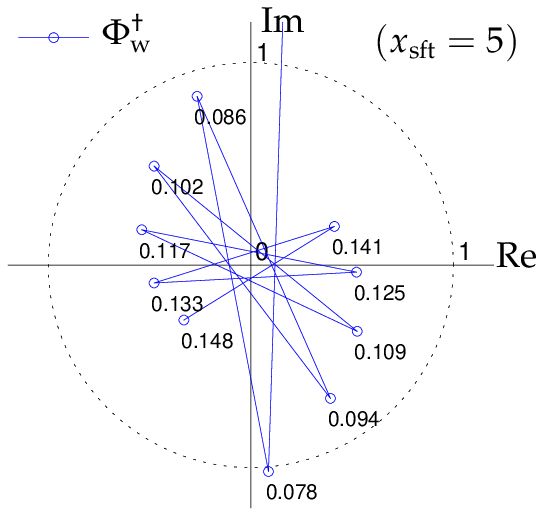}
      \hfill\Null}
    }%
  }%
  }\hfill\Null}%
 }%
 \hfill
 \parbox{0.45\hsize}{%
  {\hfill\scalebox{1.0}{\parbox{\hsize}{%
    \parbox{\hsize}{(b)\\
      {\hfill
        \includegraphics[width=\hsize]{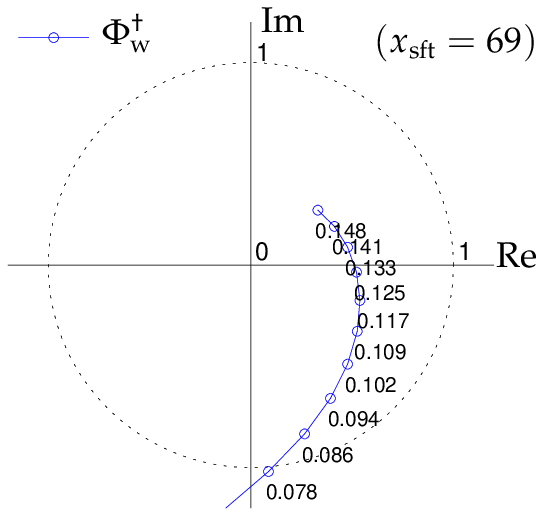}
      \hfill\Null}
    }%
  }%
  }\hfill\Null}%
 }%
 \hfill\Null
}
\hfill\Null
  \caption{
    The trajectory of the origin-shifted spectrum.
    The spectrum prior to the origin shift was evaluated by $\Phiw(\kappa)=\FT{w(x)\phi(x)}$, where $\phi(x)$ is the wavefront shown in \Fig{fig:test_sig1}(a).
    Origin-shifted spectrum $\PhiShiftw(\kappa)$ is obtained using \EQ{eq:Origin_shifted_Phi} for (a) $\xshift=5$ and (b) $\xshift=69$.
    Numbers adjacent to points indicate corresponding frequency $\kappa$.
  }%
  \label{fig:test_sig1-phase_rot_spectrum}
\end{figure}

The components of the shear frequencies are inevitably lost
during the measurement process,
as demonstrated in Sec.~\ref{subsec:shearing_transfer_function}. 
This issue can be addressed by incorporating additional difference measurements with different shear amounts
\cite{Elster:99,Elster:99-B,Elster:00,Dubra:04,Guo:12,Zhai:17-high,ling2017quadriwave}%
.
Nevertheless, this approach increases the complexity of the measurement system.
To restore these lost components using a single wavefront difference,
we employ spectral interpolation.

The interpolation technique is applied to the spectrum $\Phi(\kappa)$ before applying the inverse Fourier transform, as shown in \EQ{eq:FTinvPhiNat}, where
\begin{align}
  \Phi(\kappa)=\frac{\FT{f(x)}}{D(\kappa)}.
\end{align}
In this equation, when $\kappa_m\ne p\kappa_s$, where $p$ is any integer,
both the numerator and denominator on the right-hand side are already known,
allowing the left-hand side to be evaluated.
However, when $\kappa_m=p\kappa_s$, the spectrum needs to be interpolated from the spectra at two adjacent frequencies.

In a simple interpolation approach,
the interpolated spectrum is computed as the complex plane average of the two adjacent spectral points \cite{Liang:06}, denoted as
\begin{align}
  \InterpolateAve{\Phi_m}=\left\{
   \begin{array}{ll}
     \overline{\Phi_m}=(\Phi_{m-1}+\Phi_{m+1})/2.
      & \mbox{for }\kappa_m=p\kappa_s,
      \\
     \Phi_m 
      & \mbox{otherwise}.
   \end{array}\right.
\end{align}
Nonetheless, $\overline{\Phi_m}$ introduces a significant error owing to the rapid rotation of its complex argument.

The Fourier transform, as depicted in \EQ{eq:DFT}, involves rearranging the equation by isolating $e^{-2\pi i\kappa_m x_0}$ outside the summation, denoted as
\begin{align}
  \label{eq:DFT-dep_x0}
    A_m
      &=\sum_{n=0}^{N-1} a_n\,e^{-2\pi i\kappa_m x_n}
      =\NondepOrigin{A}_m\,e^{-2\pi i\kappa_m x_0},
    \\
  \label{eq:DFT-indep_x0}
    \NondepOrigin{A}_m
      &=\sum_{n=0}^{N-1} a_n\,e^{-2\pi i\kappa_m(x_n-x_0)}
       =\sum_{n=0}^{N-1} a_n\,e^{-i\frac{2\pi mn}{N}}.
\end{align}
In this notation, $\NondepOrigin{A}_m$ remains independent of the origin, $x_0$.
Equation (\ref{eq:DFT-dep_x0}) signifies that the complex argument of $A_m$ is the sum of the argument of $\NondepOrigin{A}_m$
and $-2\pi \kappa_m x_0$.
When the origin is shifted to
\begin{align}
  \ShiftOrigin{x}=x-\xshift,
\end{align}
the absolute value of the spectrum remains unchanged; however,
the trajectory of the spectrum in the complex plane undergoes a transformation.
Figure \ref{fig:test_sig1-phase_rot_spectrum} shows typical examples of these trajectories.
The trajectory of the spectrum in the complex plane, as shown in \Fig{fig:test_sig1-phase_rot_spectrum}(a),
exhibits a zigzag pattern owing to rapid changes in the rotation of the complex argument of the spectrum.
In contrast, the trajectory in (b) changes smoothly.
The smooth variation of the trajectory offers higher accuracy for spectral interpolation at shear harmonic frequencies.
The interpolation error is smaller in case (b) compared to case (a).

To achieve a smooth variation in the spectrum, an origin shift of $\Phi_m$ is applied prior to the spectral interpolation,
and the origin is restored after the interpolation process.
The origin-shifted spectrum is represented as
\begin{align}
  \label{eq:Origin_shifted_Phi}
  \PhiShift_m=\Phi_m e^{2\pi i \kappa_m \xshift}.
\end{align}
In this equation, the origin shift $\xshift$ is determined using the following algorithm to ensure a gradual change in the complex argument of $\PhiShift_m$.
\begin{enumerate}
  \renewcommand{\labelenumi}{Step \arabic{enumi}:}\setlength{\leftskip}{2em}
  \item Evaluate the complex argument, $\theta_m=\angle\Phi_m$ for $\kappa_m\ne p\kappa_s$.
        The results are wrapped phases; i.e., $-1/2\le\theta_m<1/2$ cycle.
  \item Unwrap $\theta_m$ as $\Unwrapped{\theta_m}$ so that the absolute difference of the result between each pair of adjacent points does not exceed a half-cycle.
  \item Evaluate $\xshift$ as a negative-averaged gradient of $\Unwrapped{\theta_m}$ by
   \begin{align}
     \xshift=-\frac{\Unwrapped{\theta_{\Max{m}}}-\Unwrapped{\theta_{\Min{m}}}}{\kappa_{\Max{m}}-\kappa_{\Min{m}}}.
   \end{align}
\end{enumerate}
Once the origin shift $\xshift$ is acquired, the origin-shifted complex argument $\thetaShift_m$
and the origin-shifted spectrum $\PhiShift_m$ can be evaluated as
\begin{align}
  \thetaShift_m&=\Unwrapped{\theta_m}+\kappa_m \xshift,\\
  \label{eq:PhiShift_m}
  \PhiShift_m&=\Abs{\Phi_m} e^{2\pi i \thetaShift_m}.
\end{align}
In this equation, $\thetaShift_m$
must be an unwrapped complex argument without any discontinuity larger than a half-cycle.
However, if the change in the argument for $\kappa_m\xshift$ is significant, $\thetaShift_m$ may have jumps that exceed a half-cycle.
In such situations, an additional origin shift is evaluated for the shifted $\PhiShift_m$ using the same procedure,
while $\Phi_m$ in step 1 is substituted by $\PhiShift_m$ in \EQ{eq:PhiShift_m},
and the resulting $\xshift$ is accumulated with the previous value.

The interpolation of the origin-shifted spectrum $\PhiIpPolarShiftSub{m}$ is evaluated by computing both the radial and angular averages
of $\PhiShift_{m-1}$ and $\PhiShift_{m+1}$ as
\begin{align}
  \label{eq:polar-interpolation0}
  \PhiIpPolarShiftSub{m}
   &=
    \overline{\Abs{\Phi_m}}\,e^{2\pi i\,\overline{\thetaShift_m}},
\end{align}
where
\begin{align}
  \label{eq:polar-interpolation-abs0}
  \overline{\Abs{\Phi_m}}
    &=\frac{1}{2}\left(\Abs{\Phi_{m-1}}+\Abs{\Phi_{m+1}}\right),
  \\
  \label{eq:polar-interpolation-theta0}
  \overline{\thetaShift_m}
    &=\frac{1}{2}\left(\thetaShift_{m-1}+\thetaShift_{m+1}\right).
\end{align}
The interpolated spectrum in the original domain is
represented as
\begin{align}
 \label{eq:polar-interpolation}
 \Interpolate{\Phi_m}\Define
 \PhiIpPolarSub{m}=\left\{
    \begin{array}{ll}
     \PhiIpPolarShiftSub{m}\,e^{-2\pi i \kappa_m \xshift} & \mbox{for\ }\kappa_m=p\kappa_s,\\
     \Phi_{m}\ & \mbox{otherwise}.
    \end{array}\right.
\end{align}

Spectrum interpolation allows us to determine all spectral components $\Phi_m$, including the zero-frequency component $\Phi_0$.
However, $\Phi_0$, which equals the dc component (the average of $\phi(x)$) and is referred to as the `piston term,'
is inherently inaccurate
considering the piston term is inevitably lost during the measurement process, regardless of the shear amount $s$.
In other words, while it can be calculated as the average of $\Phi_{-1}$ and $\Phi_{+1}$ in the spectrum interpolation, it is not significant in the experimental measurements, and hence, cannot be determined reliably.
Nonetheless, because the piston term is simply uniformly added to $\phi(x)$, it is inconsequential in interferometric measurements that focus on wavefront distortion.


\section{Numerical Simulations}
\label{sec:result}
\begin{figure}[!t]
{%
 \hfill
 \parbox[t]{\HsizeInTwocol}{%
  {\hfill\scalebox{0.9}{\parbox{\hsize}{%
   \parbox{\hsize}{%
    (a)\\
      \includegraphics[width=\hsize]{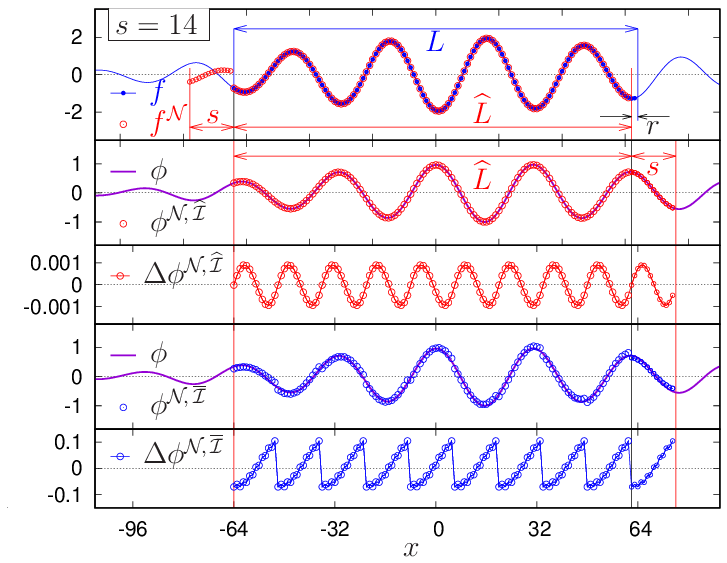}
   }%
   \\
   \parbox{\hsize}{%
    (b)\\
      \includegraphics[width=\hsize]{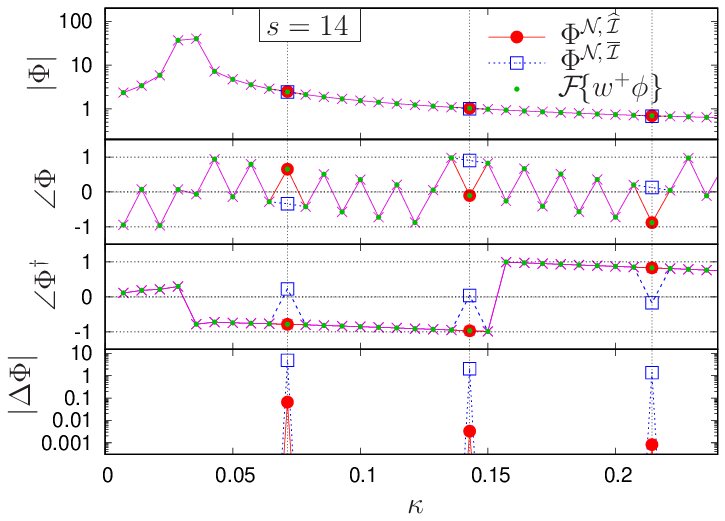}
   }%
  }}\hfill\Null}%
 }\hfill\Null%
}%
  \caption{
  Restored wavefront obtained through spectral interpolation of shearing data.
  (a)
    Analytically derived wavefront difference $f(x)$ for $s=14$ from $\phi(x)$, where $\phi(x)$ is shown in \Fig{fig:test_sig1}.
    Extended data $f^{\NatOpSym\!}(x)\equiv\NatExt{f(x)}$  using natural extension through a three-step process:
    selection of observed domain 
    and sampling,
    truncation of sampled data to ensure 
    an integer multiple of $s/\Delta x$ points, 
    and natural extension (indicated by small circles on the negative side of $\OmegaObs$, denoted as $s$). 
    Restored wavefronts for $\Phi^{\NI}(\kappa)$ and $\Phi^{\NIave}(\kappa)$ are represented
    as $\phi^{\NI}(x)$ and $\phi^{\NIave}(x)$, respectively.
  (b)
    Interpolated spectrum with and without origin shift, denoted as  $\Phi^{\NI}(\kappa)$ and $\Phi^{\NIave}(\kappa)$, respectively.
    The differences between them are only observed at $\kappa=p\kappa_s$,
    indicated by vertical dashed lines.
    The common spectrum at $\kappa\ne p\kappa_s$ is represented by cross symbols.
    The phase of $\Phi(\kappa)$ in the original domain and the origin-shifted domain ($\xshift=73.98$)
    is denoted as $\angle\Phi(\kappa)$ and
    $\angle\Phi^\dagger(\kappa)$, respectively.
    True spectrum is denoted as $\FT{w^+(x)\phi(x)}$.
    Quantities marked with $\Delta$ indicate differences from the true values.
  }
  \label{fig:test_sig1-nat_tr-s14}
\end{figure}

\begin{figure}[t]
{%
 \hfill
 \parbox[t]{\HsizeInTwocol}{%
  {\hfill\scalebox{0.9}{\parbox{\hsize}{%
   \parbox{\hsize}{%
    (a)\\
      \includegraphics[width=\hsize]{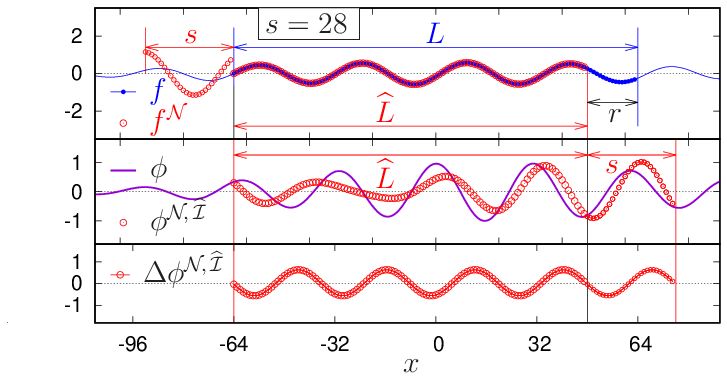}
   }%
   \\
   \parbox{\hsize}{%
    (b)\\
      \includegraphics[width=\hsize]{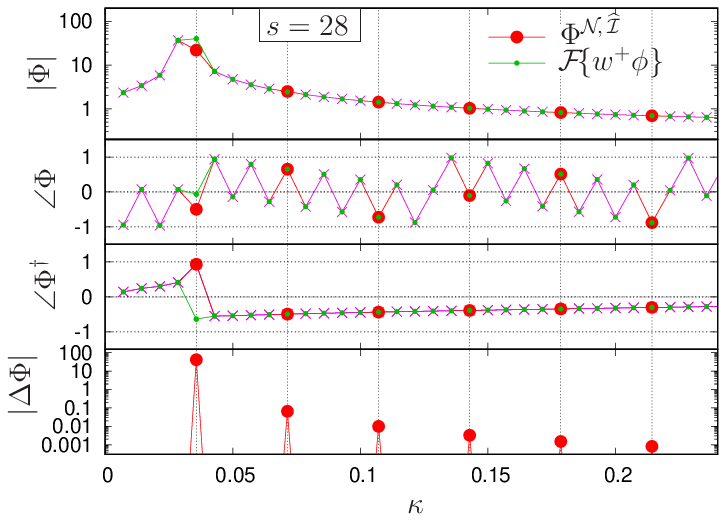}
   }%
  }}\hfill\Null}%
 }\hfill\Null%
}%
  \caption{
    Example of unsuccessful wavefront restoration.
    The symbols correspond to those in \Fig{fig:test_sig1-nat_tr-s14}.
    The difference from the example in  \Fig{fig:test_sig1-nat_tr-s14} is only the shear amount $s$; i.e., $s=28$ in this example.
    In this particular case, the shear frequency $\kappa_s$ is situated within the primary lobe of the wavefront spectrum.
  }
  \label{fig:test_sig1-nat_tr-s28}
\end{figure}

\begin{figure}[!t]
{%
 \hfill
 \parbox[t]{0.92\HsizeInTwocol}{%
      \includegraphics[width=\hsize]{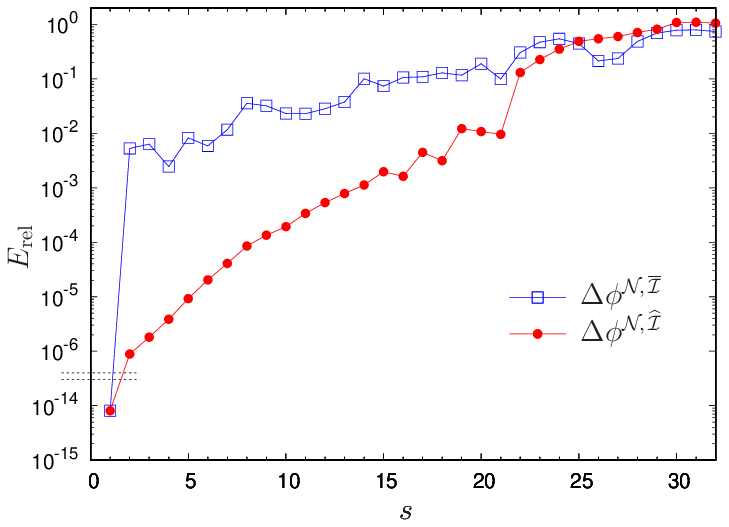}
 }%
 \hfill\Null%
}
  \caption{%
    Comparison of shear dependencies in restored wavefront error between spectral interpolation with and without origin shift.
    The vertical axis represents the normalized error, calculated as the root mean square of the true wavefront.
  }%
  \label{fig:test_sig-nat_tr-err}
\end{figure}

\begin{figure}[tb]
{%
 \hfill
 \begin{tabular}{lll}
   (a) $f_x(x,y)$
     &
   (b) $f_y(x,y)$
     &
   (c) $\phi(x,y)$
   \\
      \includegraphics[scale=0.25]{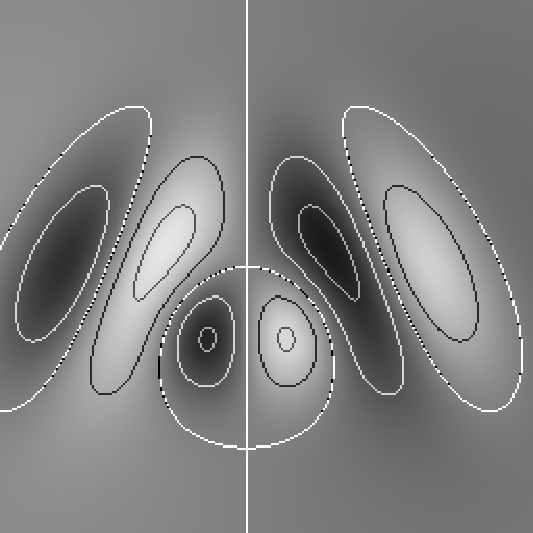}
        &
      \includegraphics[scale=0.25]{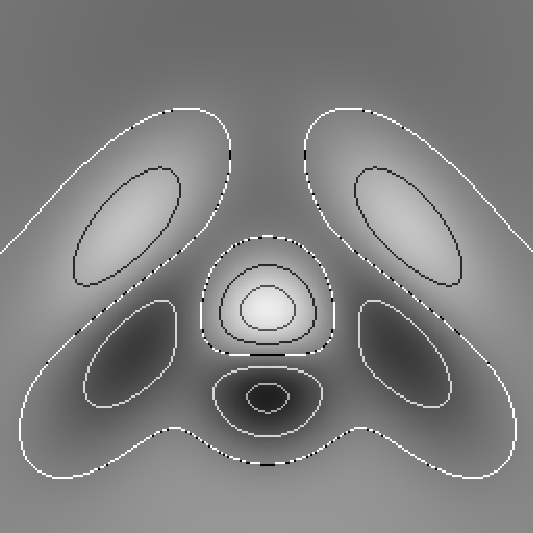}
        &
      \includegraphics[scale=0.25]{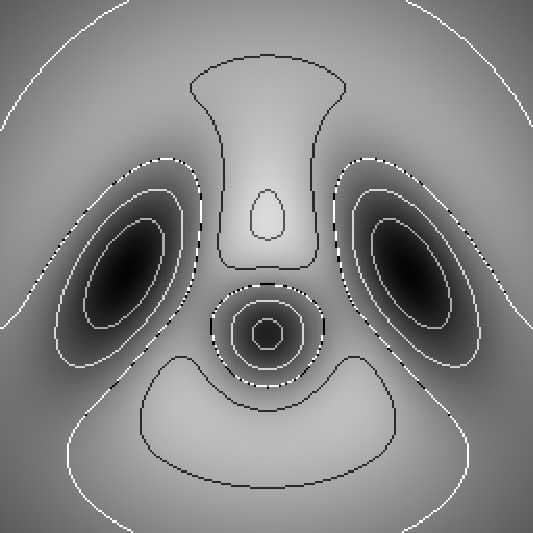}
   \\
   (d) ${\phi}_x^{\NI}(x,y)$
     &
   (e) ${\phi}_y^{\NI}(x,y)$
     &
   (f) $\phi^{\NI\dagger}(x,y)$
   \\
      \includegraphics[scale=0.25]{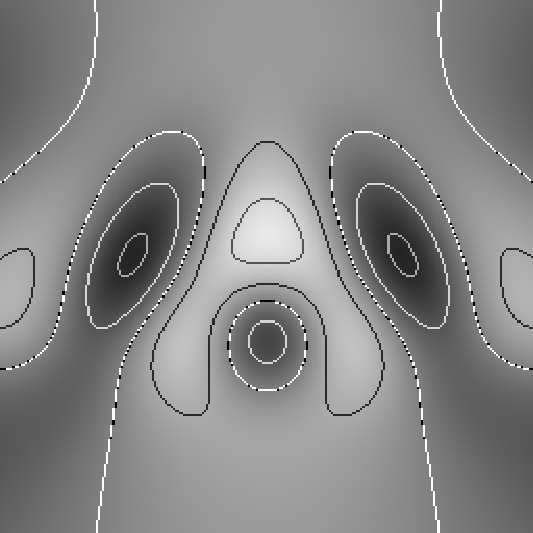}
     &
      \includegraphics[scale=0.25]{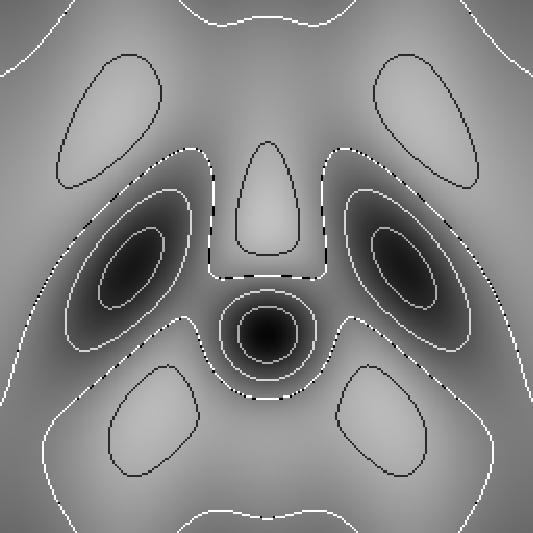}
     &
      \includegraphics[scale=0.25]{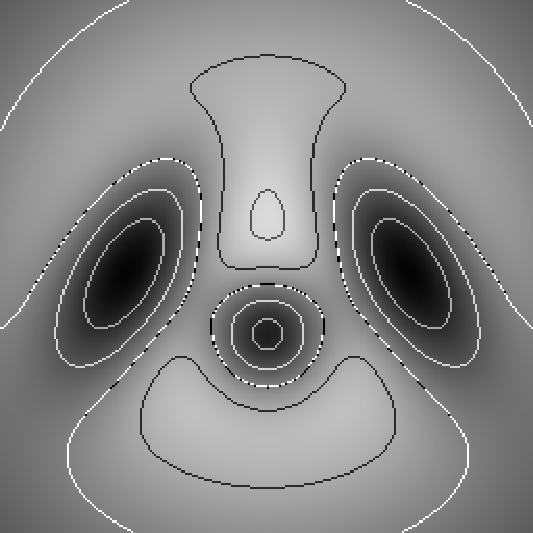}
   \\
 \end{tabular}
 \hfill\Null%
}
  \caption{Example of 2-D wavefront restoration.
    (a) and (b) show the 2-D difference functions using $\Vs=(20,0)$ and $(0,20)$ for $\phi(x,y)$, respectively,
    (c) displays the true wavefront superimposed with four Gaussian functions with elliptical contour lines,
    (d) and (e) present the restored wavefront, where $\phi_d^{\NI}(x,y)$ ($d$: either $x$ or $y$) is the restored wavefront obtained from $f_d(x,y)$.
        Each wavefront is represented by the sum of ac and dc components as $\phi_d^{\NI}(x,y)=\AC{\phi}_d^{\NI}(x,y)+\DC{\phi}$
        for comparison purposes with other wavefronts,
        where 
          $\AC{\phi}_d^{\NI}(x,y)$ represents
          a zero-biased wavefront obtained through natural extension
          combined with  spectral interpolation
          using origin shift,
          and $\DC{\phi}$ corresponds to the dc component of $\phi(x,y)$;
    (f) illustrates
        the estimated wavefront after correcting piston terms using the method proposed by Tian {\it et al}.
    Both axes in all images are denoted in $[-128,127]$.
    The brightness levels for (a) and (b) are indicated in $[-0.6,+0.6]$, while (c)-(f) are represented in $[-0.2,1.0]$.
    The contour lines have intervals of $0.2$ in all images.
  }
  \label{fig:multi_gauss-2d-nat_tr}
\end{figure}

Figure \ref{fig:test_sig1-nat_tr-s14} shows a simulated example of wavefront restoration from a wavefront difference for $s=14$ as
an exemplary case.
The observed domain size $L$ of the wavefront difference within $\OmegaObs$ is
a non-integer multiple of $s$ ($L=128$)
and requires truncation of the data to ensure $\widehat{L}$ is a multiple of $s$ ($\widehat{L}=126<L$)
to apply the natural extension technique.
After truncation, the natural extended data are added, resulting in $f^{\NatOpSym}\!(x)$.
The Fourier spectrum of the wavefront is then evaluated using $\Phi^{\NatOpSym}(\kappa)=\FT{f^{\NatOpSym}(x)}/D(\kappa)$,
which exhibits spectral leakage owing to the non-periodicity of $\phi(x)$.
In this particular example, all shear harmonic frequencies are present in the spectral leakage region.
A comparison is made between spectral interpolation with and without the origin shift,
denoted by additional superscripts $\IpPolarOpSym$ and $\IpAveOpSym$, respectively.

The absolute spectra of $\Phi^{\NI}(\kappa)$ and $\Phi^{\NIave}(\kappa)$ are nearly identical,
except for phase discrepancies at $\kappa=p\kappa_s$.
The error in the restored wavefront exhibits $s$-periodicity, as evident from $|\Delta\Phi(\kappa)|$.
This error arises solely from spectral interpolations at shear harmonic frequencies.
Furthermore, the restored wavefront exhibits an ambiguity of uniform offset
as the piston term corresponding to the spectrum at $\kappa=0$ cannot be determined,
as discussed in Sec.~\ref{subsec:interpolation}.
To facilitate a comparison between the restored and true wavefronts,
the piston term of the restored wavefront is determined using the following equation:
\begin{align}
  \phi^{\NIrep}(x)&
    =\AC{\phi}^{\NIrep}(x)+\AveObs{\Bigl.\phi(x)\Bigr.}{},
    \\
  \label{eq:phi_ac}
  \AC{\phi}^{\NIrep}(x)&
    =\FTinv{\Phi^{\NIrep}(\kappa)}-\AveObs{\FTinv{\Phi^{\NIrep}(\kappa)}}{},
\end{align}
where $\AveObs{\ }$ represents
the average of the samples in $\OmegaObs$, and the superscripts $\IpOpSym$ can be either $\IpPolarOpSym$ or $\IpAveOpSym$.

The spectral error of $\Phi^{\NI}(\kappa)$ is approximately two orders of magnitude smaller
than that of $\Phi^{\NIave}(\kappa)$.
Consequently,
when using spectral interpolation with origin shift, $\Delta\phi^{\NI}(x)$,
the error of the restored wavefront in the original spatial domain
is significantly reduced compared to the interpolation without origin shift, $\Delta\phi^{\NIave}(x)$.
Additionally, $\phi^{\NIave}(x)$ exhibits periodic discontinuities with a period $s$.
In general, it is challenging to discern whether these discontinuities arise from errors or physical phenomena.
Therefore, in wavefront restoration, 
it is crucial to achieve lower error levels while eliminating any discontinuities.

Figure \ref{fig:test_sig1-nat_tr-s28} illustrates another example for $s=28$.
In this case, the shear frequency ($\kappa_s=1/28\simeq0.0357$) is located
within the main lobe of the unlimited wavefront spectrum,
while the shear harmonic frequencies are found in the spectral tail.
In contrast to the results shown in \Fig{fig:test_sig1-nat_tr-s14} for $s=14$,
the estimation, in this case, exhibits a large error
due to $\angle\Phi^\dagger(\kappa)$ at $\kappa=\kappa_s$.
The peak and spectral width of the main lobe, $|\Phi(\kappa)|$, are estimated based on the parameters depicted in \Fig{fig:test_sig1},
showing a peak at $\kappa_{\rm pk}=1/\lambda\simeq0.0323$ and a decrease in $1/e\simeq0.37$ at $\kappa_{\rm pk}\pm\delta\kappa$,
where $\delta\kappa=1/(\pi w_g)\simeq0.0039$.
The shear frequency exhibits a significant spectral intensity component, $|\Phi(\kappa_s)|/|\Phi(\kappa_{\rm pk})|\simeq 0.42$.
At this frequency, two types of errors arise.
First, the phase changes more rapidly around the main lobe.
Second, the interpolation of intensity, as described by \EQ{eq:polar-interpolation-abs0}, introduces another source of error.
The interpolated intensity $\Abs{\PhiIpPolarSub{m}}$ is evaluated by averaging the adjacent spectral intensities;
therefore, $\Abs{\PhiIpPolarSub{m}}$ is the value between $\Abs{\PhiSub{m-1}}$ and $\Abs{\PhiSub{m+1}}$.
This implies that spectral interpolation is suitable for cases where
the spectral intensity follows a monotonically decreasing or increasing function.
Conversely, when the frequency to be interpolated is located around peaks of spectral intensity distribution, significant errors may occur.
In \Fig{fig:test_sig1-nat_tr-s28},
the actual spectrum displayed
$\Phi^{\NI}(\kappa_s)\simeq 0.3-22.1i\simeq|22.1|e^{-1.56i}$,
while $\FT{w^+\phi}_{\kappa=\kappa_s}\simeq39.4-8.1i\simeq|40.3|e^{-0.20i}$;
there are substantial disparities in both intensity and phase,
which hinders the application of spectral interpolation.
Therefore, it is necessary to choose a small shear amount $s$
that avoids the inclusion of shear harmonic frequencies within the spectral main lobe to accurately restore wavefronts.
Fortunately, selecting an appropriate $s$ is a simple task; for instance,
it can be achieved by adjusting the angle of a parallel plate relative to the optical axis in Marty's configuration.

Figure \ref{fig:test_sig-nat_tr-err} illustrates the error dependencies of $s$ by comparing the spectral interpolation methods
with and without the origin shift.
The error is evaluated using a relative error metric,
which calculates the ratio of the root mean square error of the restored wavefront to the root mean square of the bias-shifted true wavefront:
\begin{align}
  \label{eq:relative_error}
  E_{\rm rel}=\sqrt{\frac{\AveObs{\left(\phi^{\NIrep}(x)-\phi(x)\right)^2}}{\AveObs{\left(\phi(x)-\AveObs{\phi(x)}\right)^2}}}.
\end{align}
The error increases with $s$, both with and without the origin shift.
This phenomenon can be better understood by comparing the restored wavefronts in Figures \ref{fig:test_sig1-nat_tr-s14} and \ref{fig:test_sig1-nat_tr-s28}.
As $s$ increases, $k_s$ decreases and is included in the main lobe in the spectral domain.
In addition,
the extremely small error at $s=1$ is owing to the Nyquist frequency $\kappa_{\max}$; i.e., $\kappa_s=1/s>\kappa_{\max}$,
which makes spectral interpolation essentially unnecessary.

When reconstructing a 1-D wavefront, a certain ambiguity arises concerning the piston term.
As a result, when reconstructing a 2-D wavefront from a single 2-D difference function
based on the 1-D wavefront reconstruction method,
no inherent information exists to establish the relationships between piston terms across various lines.
To address such ambiguity, an additional 2-D difference function with a different shear direction is introduced.
Figure \ref{fig:multi_gauss-2d-nat_tr} illustrates a 2-D wavefront restored using two 2-D difference functions, $f_x(x,y)$ and $f_y(x,y)$, with a shear amount of 20.
These differences are computed from a true wavefront $\phi(x,y)$,
consisting of the sum of four Gaussian functions featuring elliptical contour lines defined by
$A e^{-\left(\xi^2/w_\xi^2+\eta^2/w_\eta^2\right)}$
where $\xi=(x-x_c)\cos\theta+(y-y_c)\sin\theta$ and $\eta=-(x-x_c)\sin\theta+(y-y_c)\cos\theta$.
The parameters of the Gaussian functions are $(A,x_c,y_c,w_\xi,w_\eta,\theta)=(1,0,0,150,150,0)$, $(-1,0,-32,30,30,0)$, $(-1,\pm64,0,30,60,\pm\pi/6)$.
The restored wavefront, denoted as $\AC{\phi}_x^{\NI}(x,y)$, is obtained through 1-D restoration from $f_x(x,y)$ for each $y$,
whereas the piston term is tuned using \EQ{eq:phi_ac},
ensuring that the piston terms of $\AC{\phi}_x^{\NI}(x,y)$ are set to zero for all $y$.
Similarly, the restored wavefront $\AC{\phi}_y(x,y)$ obtained from $f_y(x,y)$, exhibits zero piston terms across all $x$.

Figures \ref{fig:multi_gauss-2d-nat_tr} (d) and (e) present the restored wavefronts with a uniform bias of $\DC{\phi}=\langle{\phi(x,y)}\rangle_{w_x,w_y}$ for comparison.
On comparing $\phi_x^{\NI}(x,y)$ and $\phi_y^{\NI}(x,y)$, clear differences were noted in the restored wavefronts.
Tian, Itoh, and Yatagai \cite{Tian:95} proposed a method to determine the piston terms for $x$ and $y$ directions using a least-squares approach.
In this method, the estimated wavefront model is given as
\begin{align}
  \label{eq:Model-Tian}
  \AC{\phi}^{\dagger}(x_i,y_j)
    &=\AC{\phi}_x(x_i,y_j)+\DC{\phi}_x(y_j)+\epsilon_x(x_i,y_j)
    \nonumber\\&
     =\AC{\phi}_y(x_i,y_j)+\DC{\phi}_y(x_i)+\epsilon_y(x_i,y_j),
\end{align}
where $\AC{\phi}^{\dagger}(x_i,y_j)$ represents the wavefront to be estimated,
$\AC{\phi}_x(x_i,y_j)$ and $\AC{\phi}_y(x_i,y_j)$ are tentatively estimated wavefronts
obtained through another method,
and $\bar{\phi}_x(y_j)$ and $\bar{\phi}_y(x_i)$ are the piston terms to be determined,
while $\epsilon_x(x_i,y_j)$ and $\epsilon_y(x_i,y_j)$ represent the residuals.
The normal equations are derived to minimize $\epsilon_x^2(x_i,y_j)+\epsilon_y^2(x_i,y_j)$.
If the number of samples for $x$ and $y$ directions is $N_x$ and $N_y$, respectively,
both the number of normal equations and unknown variables ($\DC{\phi}_x(y_j)$ and $\DC{\phi}_y(x_i)$)
are equal to $N_x+N_y$.
However, Tian {\it et al}. pointed out that the normal equations are rank deficient with a rank of $N_x+N_y-1$,
which indicates that we cannot determine a 2-D piston term corresponding to a uniform offset of the 2-D wavefront.
To address this issue, we replaced one of the normal equations with the following equation:
\begin{align}
  \label{eq:2D-piston_cond}
  \langle{\DC{\phi}_x(y_j)}\rangle_{w_y}
  +\langle{\DC{\phi}_y(x_i)}\rangle_{w_x}=0.
\end{align}
By applying this equation, the 2-D piston term of $\AC{\phi}^\dagger(x_i,y_j)$ becomes zero.
In \EQtwo{eq:Model-Tian}{eq:2D-piston_cond},
the functions of $\phi$ are substituted with $\phi^{\NI}$ when estimating the tentative wavefronts via
the coupling of natural extension with origin-shift-based spectral interpolation.
Figure (f) shows the
fixed wavefront $\AC{\phi}^{\NI\dagger}(x,y)$.
A comparison between Figures (c) and (f) showed minimal error in $\phi^{\NI\dagger}(x,y)$.
The relative error of $\phi^{\NI\dagger}(x,y)$, defined similarly to the 1-D restoration presented in \EQ{eq:relative_error}, amounts to approximately $5.5\times10^{-5}$,
while both relative errors of $\phi_x^{\NI}(x,y)$ and $\phi_y^{\NI}(x,y)$ are 0.57 and 0.41, respectively.
In this particular example, the Fourier spectrum of $\phi(x,y)$ at the shear frequency ($|\kappa|=1/s$) decreases to approximately $10^{-5}$ times the maximum spectrum,
thereby indicating the application of spectral interpolations at the shear harmonic frequencies in the spectral tail region.
As a result, accurate 2-D wavefront was successfully obtained by utilizing two 2-D wavefront differences of $f_x(x,y)$ and $f_y(x,y)$.

\section{Conclusion}
\label{sec:conclusion}

In the context of wavefront restoration from observed difference data using the Fourier transform with shear amount $s$,
two significant problems need to be addressed.
First is the loss of shear harmonic components, which correspond to spatial frequencies that are multiples of $1/s$.
Second is the limited window issue, which arises when the observed data at both ends of the window fail to converge
owing to the periodicity implied in the Fourier transform.

To overcome the limited window problem, this study applied the natural extension method proposed by Elster {\it et al}.
To ensure natural extension, wherein the size of the observed data must be an integer multiple of $s$,
we chose data truncation.
We demonstrated that although the truncated data size $L^\dagger$ is narrower than the observed domain $L$,
we can achieve accurate restoration within a domain width of $L^\dagger+s$,
which is wider than both the truncated size of $L^\dagger$ and the observed domain size of $L$.
 
To address the loss of shear harmonic components, 
previous studies employed a multi-shear measurement involving different measurements with different $s$;
for a 2-D restoration, this typically entails four measurements, two for each of $x$ and $y$.
However, the use of multiple measurements for different $s$ introduces complexity to the measurement setup.
Although the application of spectral interpolation can mitigate some of these challenges,
the accuracy diminishes when using a simple spectral interpolation.
To remedy this, this study proposed applying an origin shift before spectral interpolation.
Shifting the origin in the original spatial domain
leads to a reduction in rapid phase changes within the spectral tail, thereby providing a more accurate restoration of the wavefront.
The results of 1-D numerical simulations confirmed the efficacy of coupling spectral interpolation with the origin shift technique
when the recovered frequencies were located within the spectral tail.
Nonetheless, if the lost frequencies overlapped with the main lobe, the accuracy decreased owing to rapid changes in both intensity and phase.
This limitation can be overcome in scenarios where the wavefront of the object under investigation has a smooth profile
(e.g., when the object has a smooth boundary or is composed of gas or liquid),
by employing a small $s$, which can be easily implemented in experimental setups.
Conversely, for wavefronts containing high-frequency components extending up to Nyquist frequency
(e.g., when the object under investigation has a step-like profile),
the proposed method cannot effectively reduce restoration errors.
In this case, the complex spectrum at the lost frequency cannot be smoothly interpolated
because the absolute value of the spectrum is not a smooth function
even if the complex argument is.
Recovering an accurate wavefront for such objects will require additional difference wavefront measurements with a different shear.

Furthermore, we extended our research to demonstrate accurate 2-D wavefront restoration
using the method proposed by Tian {\it et al}.
to determine ambiguities of piston terms in two sets of 1-D wavefronts, which are acquired with different shear directions,
by utilizing a least-squares approach.

Using this combination method,
an accurate 2-D wavefront can be restored from just two sheared functions with different shear directions,
in contrast to previous studies that required four sheared functions.
In actual measurements, especially for 2-D wavefront restoration,
a fractional shear problem can arise.
The approach proposed in this study is coupled with the natural extension technique that has a limitation regarding the shear amount,
namely, the shear amount must be a multiple of the sampling interval, i.e., correspond to an integer shear.
Although using the spectral interpolation technique allows the number of 2-D difference data to be reduced to two
and eliminates the limitation mandating coprime shear amounts,
the integer shear constraint, which requires more experimental efforts, persists.
When the integer shear replaces the fractional shear, a non-negligible error is introduced.
The reduction of this error is a future task.

In conclusion, this study demonstrated that accurate restoration of 2-D wavefronts
can be achieved by employing a combined approach that involves a natural extension, spectral interpolation
with origin shifting, and the least-squares method for determining piston terms.
We believe the findings of this study can contribute to the field of wavefront restoration and provide valuable insights for practical implementations.

\vspace*{2em}%
\noindent
{\small
{\bf Funding}\ 
This work was supported by JSPS KAKENHI Grant Number JP22K04117.
\\
{\bf Disclosures}\ 
The authors declare no potential conflicts of interest.
\\
{\bf Data availability}\ 
Data underlying the results presented in this paper are not publicly available at
this time but may be obtained from the authors upon reasonable request.
}%

{\footnotesize
\bibliography{paper}
}

\end{document}